\documentclass[aps,prd,twocolumn,showpacs,superscriptaddress,groupedaddress,nofootinbib]{revtex4}  

\usepackage{graphicx}  
\usepackage{dcolumn}   
\usepackage{bm}        
\usepackage{amssymb}   
\usepackage{amsmath}
\usepackage{color}

\def\be{\begin{equation}}
\def\ee{\end{equation}}
\def\bea{\begin{eqnarray}}
\def\eea{\end{eqnarray}}

\begin{document}

\widetext
{\flushright{DESY 18-061}}
{\flushright{SI-HEP-2018-14}}
{\flushright{QEFT-2018-08}}


\title{Prospects of discovering stable double-heavy tetraquarks at a Tera-$Z$ factory}

\author{Ahmed~Ali}
\email{ahmed.ali@desy.de}
\affiliation{Deutsches Elektronen-Synchrotron DESY, D-22607 Hamburg, Germany}
\author{Alexander Ya. Parkhomenko}
\email{parkh@uniyar.ac.ru}
\affiliation{Department of Theoretical Physics, P.\,G.~Demidov Yaroslavl State University, Sovietskaya 14, 150003 Yaroslavl, Russia}
\author{Qin Qin}
\email{qin@physik.uni-siegen.de}
\affiliation{Theoretische Physik 1, Naturwissenschaftlich-Technische Fakult\"{a}t,
Universit\"{a}t Siegen, Walter-Flex-Strasse 3, D-57068 Siegen, Germany}
\author{Wei Wang}
\email{wei.wang@sjtu.edu.cn}
\affiliation{INPAC, Shanghai Key Laboratory for Particle Physics and Cosmology, MOE Key Laboratory for Particle Physics,
School of Physics and Astronomy, Shanghai Jiao Tong University, Shanghai 200240, China} 
\date{\today}

\begin{abstract}
Motivated by a number of theoretical   considerations, predicting the deeply bound double-heavy tetraquarks
$T^{\{bb\}}_{[\bar u \bar d]}$,  $T^{\{bb\}}_{[\bar u \bar s]}$ and $T^{\{bb\}}_{[\bar d \bar s]}$, 
we explore the potential of their discovery at Tera-$Z$ factories. 
Using the process $Z \to b \bar b b \bar b$,
we calculate, employing the Monte Carlo generators MadGraph5$\_$aMC@NLO and Pythia6,
the phase space configuration in which the~$b b$  pair is likely to fragment as a diquark. 
In a jet-cone, defined by an invariant mass interval $m_{bb} < M_{T^{\{bb\}}_{[\bar q \bar q']}} + \Delta M$,
the sought-after tetraquarks $T^{\{bb\}}_{[\bar q \bar q^\prime]}$ as well as the double-bottom baryons,~$\Xi_{bb}^{0,-}$,  and $\Omega_{bb}^-$, 
can be produced. Using the heavy quark--diquark symmetry, we estimate 
$\mathcal{B} (Z \to T^{\{bb\}}_{[\bar u \bar d]} + \; \bar b \bar b) = (1.2^{+1.0}_{-0.3}) \times 10^{-6}$, 
and about a half of this for the $T^{\{bb\}}_{[\bar{u}\bar{s}]}$ and $T^{\{bb\}}_{[\bar d \bar s]}$. 
We also present an estimate of their lifetimes using the heavy quark expansion, yielding $\tau(T^{\{bb\}}_{[\bar q \bar q^\prime]}) \simeq 800$~fs.
Measuring the tetraquark masses would require decays, such as $T^{\{bb\} -}_{[\bar u \bar d]} \to B^- D^- \pi^+$,
$T^{\{bb\} -}_{[\bar u \bar d]} \to J/\psi \overline K^0 B^-$, $T^{\{bb\} -}_{[\bar u \bar d]} \to J/\psi K^- \overline B^0$, 
$T^{\{bb\} -}_{[\bar u \bar s]} \to \Xi_{bc}^0 \Sigma^-$, and $T^{\{bb\} 0}_{[\bar d \bar s]} \to \Xi_{bc}^0 \bar\Sigma^0$,
with subsequent decay chains in exclusive non-leptonic final states. We estimate a couple of the decay widths and find that the product branching ratios do not exceed~$10^{-5}$. 
Hence, a good fraction of these modes will be required for a discovery of $T^{\{bb\}}_{[\bar q \bar q']}$ at a Tera-$Z$ factory.  

\end{abstract}

\maketitle

\section{Introduction} 
\label{sec:introduction} 
 
The experimental discovery of exotic, hidden-charm and hidden-beauty states, has opened a new field in hadron spectroscopy.
The exotic states, called~$X$, $Y$, $Z$ and~$P_c$, have been analysed in a number of theoretical models. They have been claimed to be hybrid quarkonia, hadron molecules,  coupled-channel or threshold effects, and
multiquark states, see~\cite{Ali:2017jda, Esposito:2016noz,  Chen:2016qju, Guo:2017jvc, Olsen:2017bmm} for recent reviews and extensive references therein. 
Their dynamics is very much an open question with the models dividing themselves approximately in two classes, those
reflecting residual QCD long-distance effects, dominated by meson exchanges, and those reflecting genuine short-distance
forces, induced by gluon exchanges.  A particular realisation of the latter class of models assumes that heavy baryons and tetraquarks may  be viewed as diquark-quark and diquark-antidiquark objects, respectively, with the diquarks having well-defined color, spin and flavor quantum numbers. 
 Indeed, if tetraquarks,  which are stable against strong and radiative decays, could be found in experiments, this would provide an irrefutable
evidence of compact diquarks as building blocks of hadronic matter. 

\begin{figure}[tb]
\begin{center} 
\includegraphics[width=0.15\textwidth]{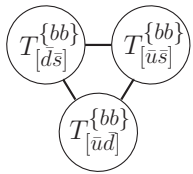} 
\end{center}
\caption{$SU(3)_F$-triplet of double-bottom tetraquark states. The spin-parity quantum numbers are $J^P = 1^+$.} 
\label{fig:doubly-bottom-tetraquark-triplet}
\end{figure}

The objects of our interest in this paper are the double-bottom $J^P = 1^+$ tetraquarks
$T^{\{bb\}}_{[\bar u \bar d]}$, and the related ones $T^{\{bb\}}_{[\bar u \bar s]}$ and $T^{\{bb\}}_{[\bar d \bar s]}$ collected in Fig.~\ref{fig:doubly-bottom-tetraquark-triplet}, 
which have evoked lately a lot of theoretical and phenomenological interest~\cite{Karliner:2017qjm, Eichten:2017ffp, Francis:2016hui, Bicudo:2017szl, Mehen:2017nrh, Czarnecki:2017vco}, 
though the possibility of stable multiquark states was already pointed out a long time ago~\cite{Ader:1981db, Manohar:1992nd}. 
Likewise, estimates of the tetraquark masses with two heavy quarks carried out in quark models well over a decade ago also predicted stable tetraquarks~\cite{Ebert:2007rn}.

Concentrating on the $T^{\{bb\}}_{[\bar u \bar d]}$ state, which is a $J^P = 1^+$, $I = 0$ tetraquark, 
consisting of the $S$-wave bound axial-vector $\{bb\}$ diquark and the scalar light $[\bar u \bar d]$ antidiquark, 
its mass is pitched at $(10389 \pm 21)$~MeV~\cite{Karliner:2017qjm}, lying about 215~MeV below the $B B^*$ threshold. 
Other estimates are somewhat higher in mass, with a $Q$-value of $-121$~MeV~\cite{Eichten:2017ffp},
$-189 \pm 10$~MeV~\cite{Francis:2016hui}, and $-60_{-38}^{+30}$~MeV~\cite{Bicudo:2017szl}. 
In all likelihood, $T^{\{bb\}}_{[\bar u \bar d]}$ is deeply bound against strong decays and will decay weakly. 
Mass estimates of the other two bound tetraquarks $T^{\{bb\}}_{[\bar u \bar s]}$ and $T^{\{bb\}}_{[\bar d \bar s]}$ likewise are below their corresponding mesonic thresholds, and hence they are also expected to decay weakly. 
Since no weakly decaying multiquark state has so far been observed, their observation would herald a new era in genuine multiquark physics.

\begin{figure}[tb] 
\begin{center}
\includegraphics[width=0.2\textwidth]{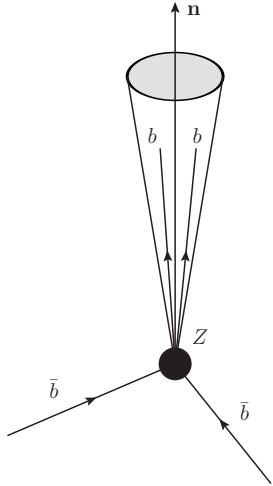}
\end{center}
\caption{Production of $b \bar b b \bar b$ in $Z$-boson decays with the $b$-quark pair forming the $bb$-jet.}
\label{fig:bbbb} 
\end{figure} 

\begin{figure}[tb]
\begin{center}
\includegraphics[width=0.15\textwidth]{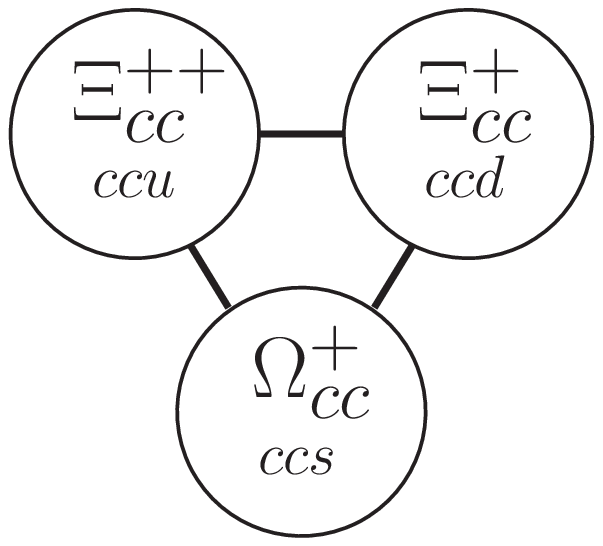}
\qquad 
\includegraphics[width=0.15\textwidth]{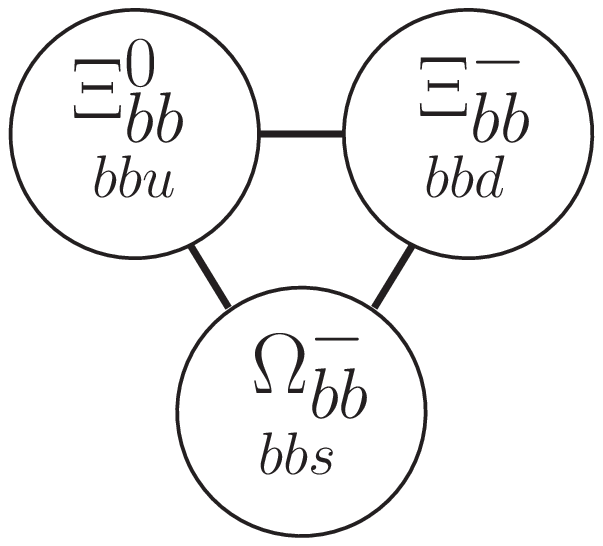}  
\end{center}
\caption{
$SU(3)_F$-triplets of double-charm (left diagram) and double-bottom (right diagram) baryons. 
Of these, the $\Xi_{cc}^{++}$ $(ccu)$ baryon has been measured by the LHCb Collab.~\cite{Aaij:2017ueg}.
} 
\label{fig:DC-DB-baryon-multiplets}
\end{figure}

At a $Z$-factory, the underlying partonic process for the searches of such tetraquarks is
\begin{equation}
Z \to b \bar{b} b \bar{b},
\label{eq:Zbbbb}
\end{equation}
which has been measured at LEP, having the branching ratio~\cite{Patrignani:2016xqp}: 
\begin{equation}
{\cal B} (Z \to b \bar b b \bar b) = \left ( 3.6 \pm 1.3 \right ) \times 10^{-4} . 
\label{eq:BRZbbbb}
\end{equation} 
The production of a double-bottom tetraquark, $T^{\{bb\}}_{[\bar q \bar q^\prime]}$, 
where the light antidiquark has the flavor content $[\bar u \bar d]$, $[\bar u \bar s]$, or $[\bar d \bar s ]$,%
\footnote{Throughout this paper, the charged conjugation is assumed.} 
is a non-perturbative process, and it involves the formation of the $bb$-diquark and its fragmentation, producing a $bb$-diquark jet. 
The conducive configuration for the diquark formation is the one where the two $b$-quarks (or two $b$-antiquarks) 
are almost collinear, satisfying $\theta_{bb} \leq \delta$, where~$\theta_{bb}$ is the angle between the $b$-quark momenta and~$\delta$ is the cone apex angle, 
and have a small relative velocity, i.\,e., their energies~$E_{b1}$ and~$E_{b2}$ differ by a small amount, $\left ( E_{b1} - E_{b2} \right) / \left ( E_{b1} + E_{b2} \right ) \leq \epsilon$, 
with $(\epsilon,\delta) \ll 1$, with an example shown in Fig.~\ref{fig:bbbb}. 
The $bb$-diquark jet can be defined, like a quark jet, by a jet resolution parameter, such as the invariant mass, or a Sterman-Weinberg jet cone~\cite{Sterman:1977wj}. 
A judicious choice of these cut-off parameters forces the two $b$-quarks to overlap in the phase space, which then fragment as a diquark.  
A tetraquark of the commensurate quark flavor is formed by picking up a light antidiquark pair $\bar u \bar d$, $\bar u \bar s$, or $\bar d \bar s$ from the debris of the jet, 
which consists of mostly soft pions or kaons. 
Likewise, the $bb$-diquark will also fragment into double-bottom baryons, $\Xi_{bb}^0 (bbu)$, $\Xi_{bb}^- (bbd)$, and $\Omega_{bb}^- (bbs)$, shown in the right-hand frame in Fig.~\ref{fig:DC-DB-baryon-multiplets}.
Outside the $bb$ jet-cone, no double-bottom hadrons (tetraquarks or double-bottom baryons) will be produced. 
Hence, the hadronic texture of the $bb$-diquark jet is anticipated to be different from the fragmentation of two $b$-quark jets, whose fragmentation products are $B$-mesons or $b$-baryons.

We calculate the branching fraction 
\begin{equation}
{\cal B} (Z \to (bb)_{\rm jet} (\Delta M) + \bar b + \bar b), 
\label{eq:BRZbbjet}
\end{equation}
by defining the $(bb)_{\rm jet}$ with a cut-off on the $bb$-pair invariant mass,~$\Delta M$.
An estimate of~$\Delta M$ can be obtained from the inclusive $B_c$-meson production in $Z$-boson decays. 
Using the inclusive cross section $\sigma (e^+ e^-\to Z \to b \bar b c \bar c)$, 
obtained via MadGraph~\cite{Alwall:2014hca} and Pythia6~\cite{Sjostrand:2006za}, 
and the inclusive $B_c$-meson cross section $\sigma (e^+ e^- \to Z \to B_c + b + \bar c)$, 
obtained by using the NRQCD-based calculations~\cite{Yang:2011ps},
one can evaluate the fraction $f (c \bar b \to B_c)$ for the $c \bar b$-pair fusion into the $B_c$ mesons%
\footnote{We would have liked to use the LEP data for this process, but it is too sparse for a quantitative analysis.}.
The phase space of the fragmentation process to produce the $B_c$-meson is limited by $m_{c \bar b} < M_\text{cut}^{c \bar b}$, 
with $M_\text{cut}^{c \bar b}$ being the maximum value of the invariant mass in which the $c \bar b$-fusion takes place. 
Beyond this cut, the $b$- and $\bar c$-quarks fragment independently. 
We use $M_\text{cut}^{c \bar b}$ to estimate~$\Delta M$, which yields us the partial branching ratio in Eq.~(\ref{eq:BRZbbjet}). 
The details of this calculation are given in Section~\ref{sec:production-Tbb}.

The final step in the calculation of the branching ratio 
\begin{equation}
{\cal B} (Z \to T^{\{bb\}}_{[\bar q \bar q^\prime]} + X), 
\label{eq:BRZTbb}
\end{equation}
is to estimate the probability of the diquark-jet $(bb)_{\rm jet} (\Delta M)$ to fragment into $T^{\{bb\}}_{[\bar q \bar q^\prime]} + X$.
There are essentially two possibilities: The $bb$-diquark jet will either fragment into double-bottom tetraquarks, shown in Fig.~\ref{fig:doubly-bottom-tetraquark-triplet}, 
or into the double-bottom baryons, shown in the right panel in Fig.~\ref{fig:DC-DB-baryon-multiplets}. 
This involves estimating the relative probability of emitting a light (anti)-diquark $[\bar u \bar d]$, $[\bar u \bar s]$, or $[\bar d \bar s]$ from the vacuum in the presence of a $bb$-diquark color source 
to that of picking up a light quark~$q$ ($q = u,\, d,\, s$) from the $q \bar q$-pair produced in the similar situation. 
The two probabilities are related by heavy quark--heavy diquark symmetry.

This probability is similar to the relative probability that a $b$-quark fragments into a $b$-baryon, such as $\Lambda_b = b u d$,
which involves picking up the diquark $[u d]$ from the vacuum, to that of a $b$-quark fragmenting into a heavy-light meson $B^- = b \bar u$ (or $\bar B^0 = b \bar d$). 
Denoting these fractions by $f_{\Lambda_b}$ and $f_{B_u} \, (f_{B_d})$, respectively, we need to know the ratio $f_{\Lambda_b}/(f_{B_u} + f_{B_d})$. 
As is well-known, this ratio has been measured differently at the Fermilab-Tevatron~\cite{Aaltonen:2008zd}: $f_{\Lambda_b}/(f_{B_u} + f_{B_d}) = 0.281 \pm 0.012^{+0.058 +0.128}_{-0.056 -0.087}$,
at the LHC by LHCb Collab.~\cite{Aaij:2011jp}, which finds the significant dependence of this ratio on~$p_T$%
\footnote{The LHCb analysis has been updated in Ref.~\cite{Aaij:2014jyk}.},
$f_{\Lambda_b}/(f_{B_u} + f_{B_d}) = (0.404 \pm 0.017 \pm 0.027 \pm 0.105) \, [1 - (0.031 \pm 0.004 \pm 0.003) \times p_T~({\rm GeV})]$,
and by LEP experiments from $Z$-boson decays $f_{\Lambda_b}/(f_{B_u} + f_{B_d}) = 0.11 \pm 0.02$~\cite{Patrignani:2016xqp}. 
The dynamical reason behind these variations is not clear, but as we are estimating the $Z$-boson branching ratios, we shall use the average from the LEP experiments.

We also present an estimate of the total lifetimes of the tetraquarks $T^{\{bb\}}_{[\bar q \bar q^\prime]}$ based on the heavy quark expansion.  
As each of the two $b$-quarks in the tetraquark will decay independently, the decay widths of $T^{\{bb\}}_{[\bar q \bar q^\prime]}$ should be approximately about twice that of the $b$-quark, 
yielding $\tau (T^{\{bb\}}_{[\bar q \bar q^\prime]}) \simeq 800$~fs. The details are given in Section~\ref{sec:lifetimes-Tbb}.

\begin{figure}[tb]
\begin{center}
\includegraphics[width=0.15\textwidth]{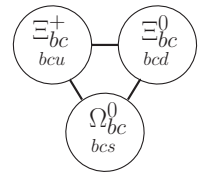} 
\end{center}
\caption{$SU(3)_F$-triplet of the bottom-charmed baryons.} 
\label{fig:BC-baryon-triplets}
\end{figure}

The tell-tale signatures of the double-bottom tetraquarks are decays into ``wrong-sign'' heavy mesons, 
such as $T^{\{bb\} -}_{[\bar u \bar d]} \to B^- D^+ \pi^-$, $\bar B^0 D^0 \pi^-$~\cite{Karliner:2017qjm, Eichten:2017ffp, Esposito:2013fma, Luo:2017eub}.
Also interesting are the three-body decays $T^{\{bb\} -}_{[\bar u \bar s]} \to B^- \left ( J/\psi, \, \psi^\prime \right ) \phi$, $B_s^0 \left ( J/\psi, \, \psi^\prime \right ) K^-$  
involving the $J/\psi$- or $\psi^\prime$-mesons, and into bottom-charmed baryons (see Fig.~\ref{fig:BC-baryon-triplets}), 
such as $T^{\{bb\} -}_{[\bar u \bar d]} \to \Xi_{bc}^0 \bar p$, $\Omega_{bc}^0 \bar\Lambda_c^-$. 
These baryons also remain to be discovered, and their discovery channels have been recently presented in Refs.~\cite{Wang:2017azm, Shi:2017dto}.  
This underscores the huge potential of the Tera-$Z$ colliders in mapping out the landscape of the double-heavy baryons and double-heavy tetraquarks.

The rest of this work is organised as follows. 
In Sec.~\ref{sec:production-Tbb}, we present an estimate of the branching ratios of double-bottom tetraquarks in $Z$-boson decays. 
Lifetimes of the tetraquarks are discussed in Sec.~\ref{sec:lifetimes-Tbb}. 
In Secs.~\ref{sec:weak-decays-Tbbud} and \ref{sec:weak-decays-Tbbus}, we discuss the discovery modes for the $T_{[\bar u\bar d]}^{\{bb\}}$, $T_{[\bar u \bar s]}^{\{bb\}}$  and  $T_{[\bar d \bar s]}^{\{bb\}}$ tetraquarks, respectively.  A special type of decays induced by $W$-exchange diagrams is discussed in Sec.~\ref{sec:W-exchange}. A brief conclusion is given in the last section.

\section{Production of double-bottom tetraquarks in $Z$-boson decays}
\label{sec:production-Tbb}

As already stated, at a $Z$-factory, double-heavy tetraquarks $T^{\{bb\}}_{[\bar q \bar q']}$ are anticipated to be produced 
in the partonic process $Z \to b \bar{b} b \bar{b}$, with the two $b$-quarks (or two $b$-antiquarks) moving collinearly and 
having a small relative velocity. A reasonable cut on the invariant mass of the two $b$-quarks  ensures these constraints. 

To calculate the branching ratio in~(\ref{eq:BRZTbb}), we invoke the decay $Z \to B_c + X$, which has been measured at LEP. 
The idea is to calculate the fraction of the $c \bar b$ quarks which hadronise into the $B_c$-mesons, $f (c \bar b \to B_c)$, 
in the underlying process $Z \to b \bar b c \bar c$.
Theoretically, the $B_c$ production cross section has been calculated at the leading order (LO) in the NRQCD framework,
yielding: $\sigma (e^+ e^- \to Z \to B_c + b + \bar c) = (5.19^{+6.22}_{-2.42})$~pb~\cite{Yang:2011ps}. 
The central (upper, lower) value corresponds to the input values of the quark masses $m_b = 4.9$~GeV and $m_c = 1.5$~GeV 
($m_b = 5.3$~GeV and $m_c = 1.2$~GeV; $m_b = 4.5$~GeV and $m_c = 1.8$~GeV). 
With the same input, we have generated three sets of 10000 LO parton showered $e^+ e^- \to b \bar b c \bar c$ events 
at the $Z$-boson mass using MadGraph~\cite{Alwall:2014hca} and Pythia6~\cite{Sjostrand:2006za}, varying the bottom and charmed quark mass values. 
(The mass parameters used by MadGraph are the pole masses. The difference between the $1S$ mass value used in~\cite{Yang:2011ps} 
and the pole mass of a heavy quark is of the $\alpha_s^2$ order (see~(1) of~\cite{Hoang:1999cj}), which can be neglected safely.) 
The cross section $\sigma (e^+ e^-\to Z \to b \bar b c \bar c)$, evaluated using MadGraph yields 64.50~pb for the central values of the quark masses 
($m_b = 4.9$~GeV, $m_c = 1.5$~GeV), 76.79~pb for the upper ($m_b = 5.3$~GeV, $m_c = 1.2$~GeV) and 56.75~pb for the lower values of the masses 
($m_b = 4.5$~GeV and $m_c = 1.8$~GeV). This yields $f (c \bar b \to B_c) = 8.05\%$, $14.86\%$ and $4.88\%$, respectively%
\footnote{The NLO MadGraph result for the cross section is enhanced by more than a factor of~2 compared with the LO result, but as the corresponding NRQCD 
calculation is done only in the LO approximation, we use the LO simulation to be consistent with Ref.~\cite{Yang:2011ps}.}. 
With the $m_{c \bar b}$ distribution of the generated events, we find that the invariant mass cuts $m_{c \bar b} < M_{B_c} + 2.7$~GeV, 
$M_{B_c} + 4.0$~GeV and $M_{B_c} + 2.2$~GeV give the correct ratios $f(c \bar b \to B_c)$ in the three cases. All the results are collected in Table~\ref{tb:Bc}.

\begin{table}[!htbh]
  \caption{Central, lower and upper values for the cross sections $\sigma (B_c b \bar c)$ and $\sigma (b \bar b c \bar c)$,
  the fraction $f (c \bar b \to B_c)$, and the invariant mass upper bound $\Delta M$ for the process $e^+ e^- \to b \bar b c \bar c$.}
  \label{tb:Bc}
  \centering
  \begin{tabular}[t]{|l|c|c|c|c|}\hline
 \parbox[0pt][2em][c]{0cm}{Quark masses}  & $\sigma(B_c  b  \bar{c})$ [pb] & $\sigma(b\bar{b}c\bar{c})$ [pb] & $f(\bar{b}c\to B_c)$ & $\Delta M$ [GeV] \\ \hline
 \parbox[0pt][2em][c]{0cm}{}central  & 5.19 & 64.50 & 8.05\% & 2.7 \\ 
 \parbox[0pt][2em][c]{0cm}{}upper  & 11.41 & 76.79 & 14.86\% & 4.0  \\ 
 \parbox[0pt][2em][c]{0cm}{}lower  & 2.77 & 56.75 & 4.88\% & 2.2  \\ 
  \hline
  \end{tabular}
\end{table}

\begin{figure}[tb] 
\begin{center}
\includegraphics[width=0.4\textwidth]{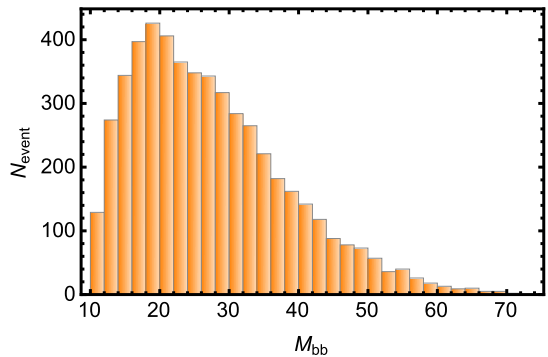} 
\end{center}
\caption{The $bb$-quark-pair invariant-mass distribution for the process $e^+ e^- \to Z \to (bb)_{\rm jet} + \bar{b} + \bar{b}$
obtained by generating $10^{4}$ events using MadGraph and Pythia6  at the  NLO accuracy. The $ (bb)_{\rm jet}$ is defined by
the interval $M _{bb} (\Delta M)$.
} 
\label{fig:invmass} 
\end{figure}

We then generate 10000 showered $e^+ e^-\to b \bar b b \bar b$ events at the $Z$-boson mass as the centre-of-mass energy with MadGraph~\cite{Alwall:2014hca} and Pythia6~\cite{Sjostrand:2006za} 
at the NLO of QCD, with the $b$-quark pair invariant mass distribution displayed in Fig.~\ref{fig:invmass}. 
With the cross sections by MadGraph, the branching ratio $\mathcal{B} (Z \to b \bar b b \bar b)$ is obtained to be  
$3.23 \times 10^{-4}$, which is consistent with the world average ($3.6 \pm 1.3) \times 10^{-4}$~\cite{Patrignani:2016xqp}. 

Next, we employ the same kinematics cuts, namely  $M_{bb} \leq M_{T^{\{bb\}}_{[\bar u \bar d]}} + 2.7$~GeV,
$M_{T^{\{bb\}}_{[\bar u \bar d]}} + 4.0$~GeV, and $M_{T^{\{bb\}}_{[\bar u \bar d]}} + 2.2$~GeV, 
and find that $4.9\%$ ($5.3\%$), $8.9\%$ ($9.2\%$) and $3.8\%$ ($3.9\%$) 
of the $e^+ e^- \to Z \to b \bar b b \bar b$ events fragment into a hadron with the $bb$ ($\bar b \bar b$) quark pair.  
The numerical differences between the charged conjugated processes reflect statistical fluctuations, 
which will disappear in simulations with higher statistics. 
We average them and list the results of the ratios $f (b b \to H_{\{bb\}})$ in Table~\ref{tb:bb}. 
\begin{table}[!htbh]
  \caption{The fraction of the events $f (b b \to H_{\{bb\}})$ in the simulation for the process $e^+ e^- \to Z \to b \bar b b \bar b$, 
           which survive the indicated invariant mass cuts on the $bb$ and $\bar b \bar b$ invariant masses.} 
  \label{tb:bb}
  \centering
  \begin{tabular}[t]{|l|c|c|c|}\hline
 \parbox[0pt][2em][c]{0cm}{} $\Delta M ({\rm GeV})$ & 2.7 & 4.0 & 2.2  \\  \hline
 \parbox[0pt][2em][c]{0cm}{} $f (b b \to H_{\{bb\}})$ & 5.1\% & 9.0\% & 3.9\% \\
   \hline
  \end{tabular}
\end{table}

Thus, our simulation yields
\begin{equation}
f (b b \to H_{\{bb\}}) =  (5.1^{+3.9}_{-1.2})\% .
\label{eq:Hbb}  
\end{equation}
The uncertainty in the fragmentation fraction is sizeable, and, within $\pm 1 \sigma$, it lies in the range $(4 - 9)$\%. 
The double-bottom hadrons~$H_{\{bb\}}$ include the double-heavy tetraquarks $T^{\{bb\}}_{[\bar q \bar q^\prime]}$ 
and the double-bottom baryons $\Xi_{bb}^0 (bbu)$, $\Xi_{bb}^- (bbd)$, and $\Omega_{bb}^- (bbs)$.

\begin{figure}[tb] 
\begin{center}
\includegraphics[width=0.3\textwidth]{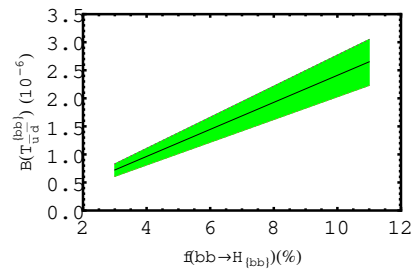} 
\end{center}
\caption{The branching ratio $\mathcal{B} (Z \to T^{\{bb\}}_{[\bar u \bar d]} + \; \bar b \bar b)$ plotted against the fraction $f (b b \to H_{\{bb\}})$.} 
\label{fig:X-section} 
\end{figure} 

The product of $\mathcal{B} (Z \to  b \bar b b \bar b)$, given in Eq.~(\ref{eq:BRZbbbb}),
$f (b b \to H_{\{bb\}})$, calculated by us and given in Eq.~(\ref{eq:Hbb}), and the fraction 
$f (b b \to T^{\{bb\}}_{[\bar u \bar d]} + X)/f (b b \to \Xi_{bb} + X)$, 
which we assume is close to $f_{\Lambda_b}/f_{B}$, derived from 
$f_{\Lambda_b}/(f_{B_u} + f_{B_d}) = 0.11 \pm 0.02$~\cite{Patrignani:2016xqp}, yields  
\begin{equation}
\mathcal{B} (Z \to T^{\{bb\}}_{[\bar u \bar d]} + \; \bar b \bar b) = (1.2^{+1.0}_{-0.3}) \times 10^{-6} .
\label{eq:Tbbcross}  
\end{equation}
We show the dependence of the branching ratio $\mathcal{B} (Z \to T^{\{bb\}}_{[\bar u \bar d]} + \; X)$ on $f (b b \to H_{\{bb\}})$ in Fig.~\ref{fig:X-section}. 
The same estimate applies to the charged conjugated process. 
We expect approximately half of this number for the production of $T^{\{bb\}}_{[\bar u \bar s]}$ and $T^{\{bb\}}_{[\bar d \bar s]}$ in $Z$-decays. 

The branching ratios for the double-bottom baryons (summed over the states) are estimated as
\begin{equation}
\mathcal{B} (Z \to (\Xi_{bb}^0, \Xi_{bb}^-, \Omega_{bb}^-) + X) : 
\mathcal{B}(Z \to  T^{\{bb\}}_{[\bar q \bar q']} + X) \approx 5.8 : 1.
\label{eq:BRXibb}
\end{equation}
Thus, we anticipate about six times as many double-bottom baryons as the double-bottom tetraquarks in the $Z$-boson decays.

\section{Lifetimes}
\label{sec:lifetimes-Tbb}

Before discussing the weak decays, we present an estimate of the lifetimes, the inverse of total decay rates. 
The decay width of $T^{\{bb\}}_{[\bar q \bar q']}$ into an inclusive final state~$X$ can be expressed as a phase space integral over the matrix element squared of the effective Hamiltonian sandwiched between the initial $| T^{\{bb\}}_{[\bar q \bar q']} \rangle$ and final $\langle X |$ states. 
Summing over all final states one has:
\begin{eqnarray}
\Gamma ( T^{\{bb\}}_{[\bar q \bar q']} \to X) = \frac{1}{2 m_T} 
\sum_{X} \int \prod_i \left [ \frac{d^3\overrightarrow{p}_i}{(2\pi)^3 2 E_i} \right ] 
\label{eq:Gamma-Tbbqqp-standard} \\
\times (2\pi)^4 \delta^{(4)} (p_T - \sum_i p_i) \, 
\frac{1}{3} \, \sum_\lambda | \langle X | {\cal H}_{\rm eff}^{\rm ew} | T^{\{bb\}}_{[\bar q \bar q']} \rangle |^2 ,  
\nonumber 
\end{eqnarray}
where $m_T$ and $p_T^\mu$ are the mass and four-momentum of the tetraquark, respectively.   
The factor~$1/3$ results after averaging over the polarization states~$\lambda$ of the tetraquark with the spin-parity $J^P = 1^+$. 
The effective electroweak Hamiltonian ${\cal H}_{\rm eff}^{\rm ew}$ is governing the weak decays to be given in the following discussions. 
Using the optical theorem, one can rewrite the total decay width as
\begin{eqnarray}
\Gamma (T^{\{bb\}}_{[\bar q \bar q']}\to X ) = \frac{1}{2 m_T} \, 
\frac{1}{3} \, \sum_\lambda \, \langle T^{\{bb\}}_{[\bar q \bar q']} | {\cal T} | T^{\{bb\}}_{[\bar q \bar q']} \rangle ,
\label{eq:Gamma-Tbbqqp-optical}
\end{eqnarray}
with the transition operator~${\cal T}$ as the time-ordered non-local double insertion of the effective Hamiltonian:
\begin{eqnarray}
{\cal T} = {\rm Im}\;\; i \int d^4 x \, T \left [ {\cal H}_{\rm eff}^{\rm ew} (x), {\cal H}_{\rm eff}^{\rm ew} (0) \right ] . 
\label{eq:Toptical}
\end{eqnarray}

The heavy-quark expansion, the operator product expansion in essence, greatly simplifies the decay widths of inclusive decays. 
Integrating over the off-shell intermediate states in Eq.~\eqref{eq:Toptical} results in a set of local operators with increasing dimensions. 
Higher dimensional operators are suppressed by inverse powers of the heavy-quark mass. 
(For a recent review see Ref.~\cite{Lenz:2014jha}.)  
Up to dimension~6, we have the transition operator expanded as~\cite{Lenz:2014jha}:
\begin{eqnarray}
{\cal T} = \frac{G_F^2 m_b^5}{192 \pi^3} \left | V_\text{CKM} \right |^2 
\left [ c_{3,b} \left (\bar b b \right ) + 
\frac{c_{5,b}}{m_b^2} \left ( \bar b g_s \sigma_{\mu\nu} G^{\mu\nu} b \right ) \right.  
\nonumber\\
+ \left . 2 \, \frac{c_{6,b}}{m_b^3} \left ( \bar b q \right )_{\Gamma} \left ( \bar q b \right )_{\Gamma} + \ldots \right ] ,
\label{eq:T-expanded} 
\end{eqnarray}
where $G_F$ is the Fermi constant and $V_\text{CKM}$ are the elements of the Cabibbo-Kobayashi-Maskawa (CKM) mixing matrix.
The coefficients~$c_{i,b}$ are the corresponding short-distance coefficients from the semileptonic and nonleptonic heavy quark decays. 
These short-distance coefficients have been precisely calculated using QCD perturbation theory~\cite{Lenz:2014jha}.

The total decay width is then determined by matrix elements of these operators, and at the leading order in~$1/m_b$, only the $\bar b b$ operator contributes: 
\begin{eqnarray}
\Gamma (T^{\{bb\}}_{[\bar q \bar q']}) = \frac{G_F^2 m_b^5}{192 \pi^3} \left | V_\text{CKM} \right |^2 
c_{3,b} \, \frac{1}{3} \, \sum_\lambda \, 
\frac{\langle T^{\{bb\}}_{[\bar q \bar q']}| \bar b b| T^{\{bb\}}_{[\bar q \bar q']} \rangle}{2 m_T} .
\label{eq:Gamma-Tbbqqp-leading}
\end{eqnarray}
The matrix element 
\begin{eqnarray}
\frac{1}{3} \, \sum_\lambda \, 
\frac{\langle T^{\{bb\}}_{[\bar q \bar q']} | \bar b b | T^{\{bb\}}_{[\bar q \bar q']} \rangle}{2 m_T}
\label{eq:ME-Tbbqqp-leading}
\end{eqnarray}
corresponds to the bottom-quark number in the $T^{\{bb\}}_{[\bar q \bar q']}$ state and is twice the matrix element for $B$-meson and $\Lambda_b$-baryon. 
Accordingly, we expect that the lifetime of the double-bottom tetraquark is one half of that of the $B$-meson~\cite{Patrignani:2016xqp}:
\begin{eqnarray}
\tau (T^{\{bb\}}_{[\bar q \bar q']}) \sim \frac{1}{2} \times 1.6 \times 10^{-12}~\text{s} = 0.8~\text{ps} . 
\label{eq:lifetime-Tbbqqp}
\end{eqnarray}
This is approximately twice the corresponding lifetime $\tau (T^{\{bb\}}_{[\bar q \bar q']}) \simeq 367$~fs estimated in~\cite{Karliner:2017qjm}.

\section{Weak Decays of $T^{\{bb\}}_{[\bar u \bar d]}$}
\label{sec:weak-decays-Tbbud}

If a state is below the lowest hadronic threshold, its decay width is determined by weak interactions. 
The decay modes are dominated by flavor-changing charged currents in the effective Hamiltonian~\cite{Manohar:2000dt}: 
\begin{eqnarray} 
{\cal H}_{\rm eff}^{(cc)} & = & \frac{4 G_F}{\sqrt 2} \, 
V_{cb} V_{ud}^* \left\{
C_1 \left [ \bar c_\alpha \gamma_\mu P_L b^\alpha \right ] 
    \left [ \bar  d_\beta \gamma^\mu P_L  u^\beta \right ] 
\right.  
\label{eq:H-eff-cc} \\ 
&& 
\left. \hspace*{11mm}
+ C_2 \left [ \bar  c_\beta \gamma_\mu P_L b^\alpha \right ]  
      \left [ \bar d_\alpha \gamma^\mu P_L  u^\beta \right ]  
\right\} 
\nonumber \\ 
& + & \frac{4 G_F}{\sqrt 2} \, 
V_{cb} V_{cs}^* \left\{ 
C_1 \left [ \bar c_\alpha \gamma_\mu P_L b^\alpha \right ] 
    \left [ \bar  s_\beta \gamma^\mu P_L  c^\beta \right ] 
\right. 
\nonumber \\ 
&& \left. \hspace*{11mm}
+ C_2 \left [ \bar  c_\beta \gamma_\mu P_L b^\alpha \right ]  
      \left [ \bar s_\alpha \gamma^\mu P_L  c^\beta \right ]  
\right\} + {\rm h.\,c.}
\nonumber 
\end{eqnarray}
Here,~$\alpha$ and~$\beta$ are 
the quark color indices, and $P_L = \left ( 1 - \gamma_5 \right )/2$ 
is the left-handed projector. The coefficients~$C_1$ and~$C_2$ are 
determined after matching the effective theory and the Standard Model, 
being scale-dependent quantities. 

We start by considering the two-body decays of $T^{\{bb\}}_{[\bar u \bar d]}$ 
and continue by taking up the three-body ones which can be divided 
into the purely mesonic and baryonic decays.

\subsection{Two-Body Baryonic Decays}   
\label{ssec:baryonic-two-body-decays}

\begin{figure}[tb] 
\begin{center}
\includegraphics[width=0.19\textwidth]{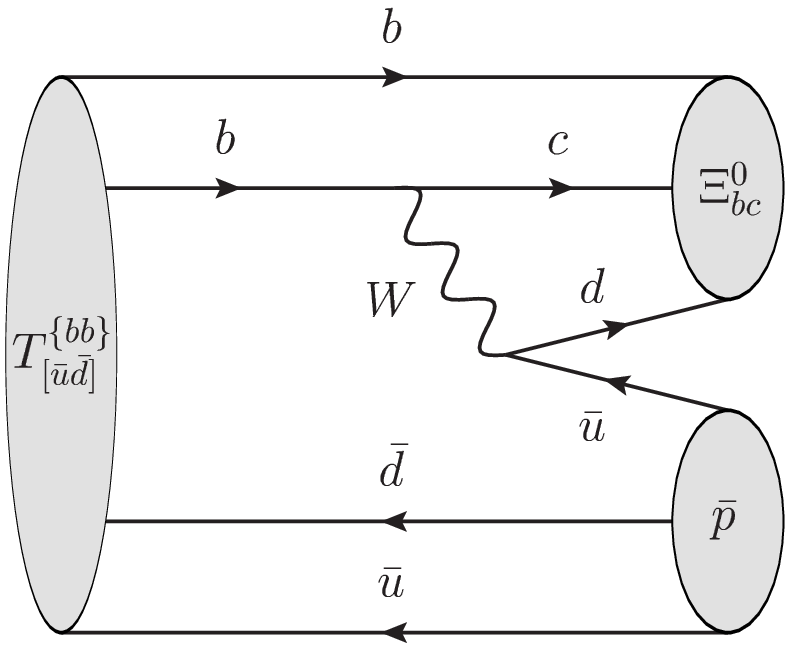} 
\qquad 
\includegraphics[width=0.19\textwidth]{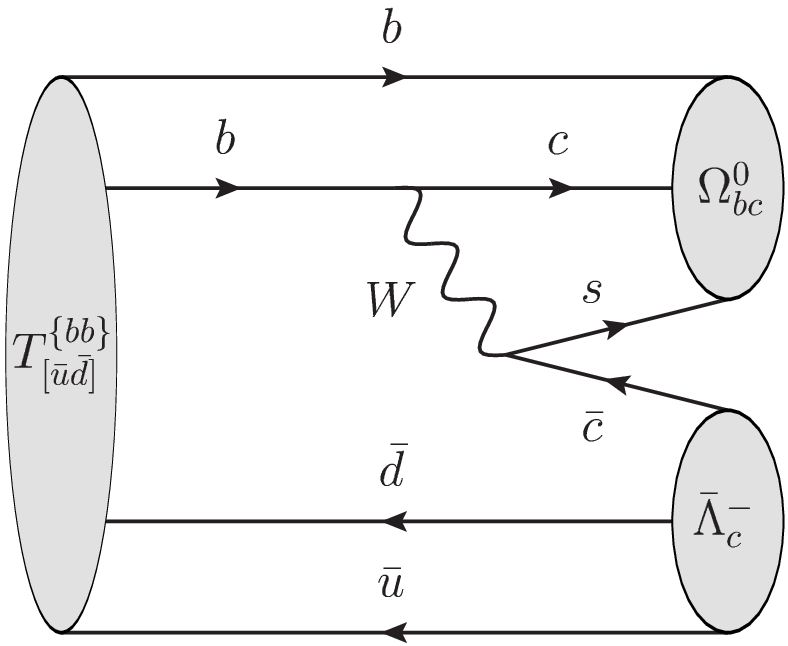}  
\end{center}
\caption{
Feynman diagrams which yield  two-body baryonic final states 
in $ T^{\{bb\} -}_{[\bar u \bar d]}$ decays 
due to the $b$-quark decay $b \to c + d + \bar u$ (left panel) 
and $b \to c + s + \bar c$ (right panel). 
} 
\label{fig:Tbbud-to-Xibc0-p-bar} 
\end{figure}

As the doubly bottom tetraquark $T^{\{bb\}}_{[\bar u \bar d]} (10482)$\footnote{
The mass assignment is according to the predictions of Ref.~\cite{Eichten:2017ffp}.} 
consists of two $b$-quarks and two antiquarks,~$\bar u$ and~$\bar d$, 
the weak decay of one of the $b$-quarks results into two possible 
Cabibbo-allowed decay channels at the quark level: $b \to c + d + \bar u$ or 
$b \to c + s + \bar c$. From three quarks and three antiquarks one can easily 
construct baryon and anti-baryon, as shown in Fig.~\ref{fig:Tbbud-to-Xibc0-p-bar}. 
Baryonic states produced in these modes have in addition to the well-known
anti-baryons, also  the so-far undiscovered  bottom-charmed baryons shown  
in Fig.~\ref{fig:BC-baryon-triplets}. They are being searched for at the LHC and
are also a part of the experimental programme at future electron-positron colliders.  
Amplitudes of these decays are non-factorisable as a quark and antiquark 
produced in the weak transition hadronise into baryon and anti-baryon, 
respectively (see Fig.~\ref{fig:Tbbud-to-Xibc0-p-bar}). Taking into account 
the axial-vector nature of the tetraquark, with $J^P = 1^+$, the general form 
of the decay amplitude induced by the $b \to c + d + \bar u$ quark channel is as follows: 
\begin{eqnarray} 
&& 
{\cal M} ( T^{\{bb\} -}_{[\bar u \bar d]} \to \Xi_{bc}^0 \, \bar p ) =  
\bar v (p_p) \left [ 
f_1^{\Xi_{bc} \bar p} \, q_\mu + 
f_2^{\Xi_{bc} \bar p} \, \gamma_\mu 
\right. \nonumber\\ 
&& \hspace*{10mm} +  
f_3^{\Xi_{bc} \bar p} \, \sigma_{\mu\nu} \, \frac{q^\nu}{m_T} +
g_1^{\Xi_{bc} \bar p} \, \gamma_5 \, q_\mu +
g_2^{\Xi_{bc} \bar p} \, \gamma_\mu \gamma_5  
\nonumber \\ 
&& \hspace*{10mm} + \left. 
g_3^{\Xi_{bc} \bar p} \, \sigma_{\mu\nu} \gamma_5 \, 
\frac{q^\nu}{m_T}    
\right ] u (p_{\Xi_{bc}}) \, \varepsilon_T^\mu (p_T) , 
\label{eq:Tbbud-to-Xibc0-p-bar} 
\end{eqnarray}
where $u (p_{\Xi_{bc}})$ and $v (p_p)$ are the wave-functions of the $\Xi_{bc}^0$-baryon 
and antiproton with the four-momenta~$p_{\Xi_{bc}}$ and~$p_p$, respectively, $q = p_{\Xi_{bc}} - p_p$, 
$\varepsilon_T^\mu (p_T)$ is the polarisation vector of the axial-vector tetraquark 
with the four-momentum $p_T = p_{\Xi_{bc}} + p_p$ and mass~$m_T$. The 
$f_i^{\Xi_{bc} \bar p}$ and $g_i^{\Xi_{bc} \bar p}$, with $i = 1,\, 2,\, 3$, 
are two sets of form factors, to be evaluated at a known kinematic point.  
Similar amplitude corresponds to the 
$T^{\{bb\} -}_{[\bar u \bar d]} \to \Omega_{bc}^0 \, \bar \Lambda_c^-$ decay 
described by the right diagram in Fig.~\ref{fig:Tbbud-to-Xibc0-p-bar}.

Inspired by the $B$-meson decay data~\cite{Patrignani:2016xqp}:  
\begin{eqnarray}
{\cal B} (\overline B^0\to D^+\pi^-) &=& (2.52 \pm 0.13) \times 10^{-3},  \nonumber\\
{\cal B} (\overline B^0\to D^+D_s^-) &=& (7.2  \pm 0.8 ) \times 10^{-3}, 
\label{eq:B-to-D-decays-exp} 
\end{eqnarray}
one infers that the branching fractions for $T^{\{bb\} -}_{[\bar u \bar d]} \to \Xi_{bc}^0 \, \bar p$ and $ T^{\{bb\} -}_{[\bar u \bar d]} \to \Omega_{bc}^0 \, \bar \Lambda_c^-$ decays might also reach the same magnitude, i.\,e., of order of~$10^{-3}$.  

One needs to further reconstruct the bottom-charmed baryons~$\Xi_{bc}^0$ and~$\Omega_{bc}^0$. 
Including the decay chain $\Xi_{bc}^0 \to \Lambda_b K^- \pi^+$, we find that the two-body baryonic decay modes 
of the double-bottom tetraquarks are expected to have branching fractions of order of~$10^{-6}$.

\subsection{
Three-body Mesonic Decay Modes of $T^{\{bb\} -}_{[\bar u \bar d]}$  
} 

\subsubsection{Open-Charm Final States} 

\begin{figure}[tb] 
\begin{center}
\includegraphics[width=0.19\textwidth]{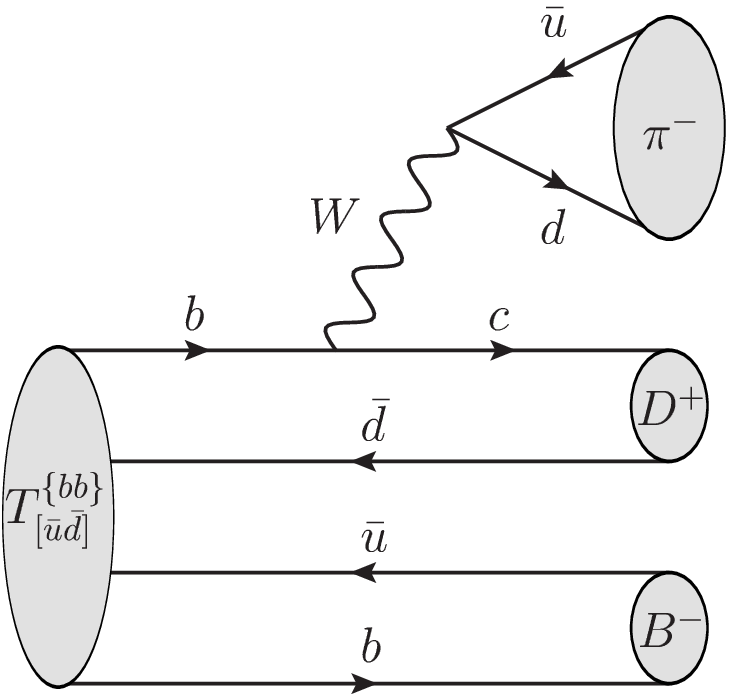}  
\qquad 
\includegraphics[width=0.19\textwidth]{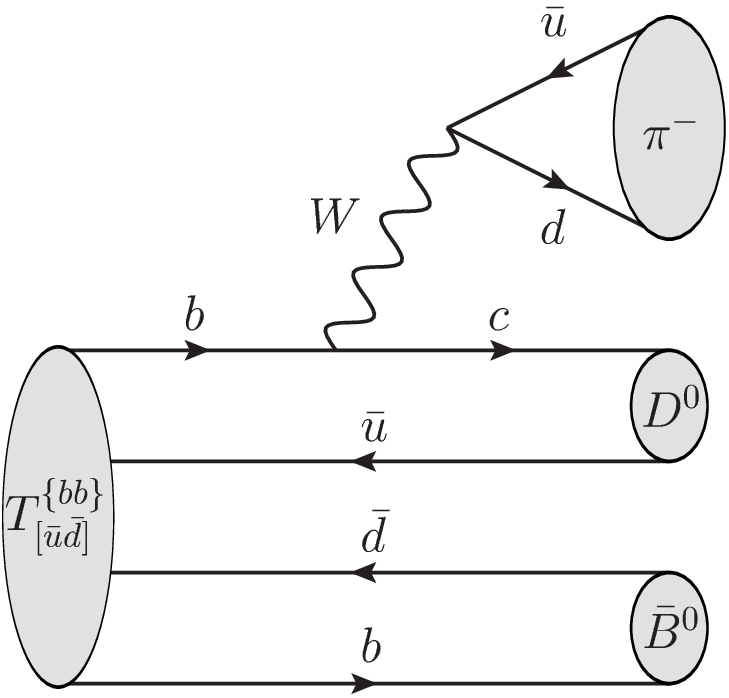}  
\end{center}
\caption{
Feynman diagrams which yield three-body mesonic final states 
in $ T^{\{bb\} -}_{[\bar u \bar d]}$ decays
due to the $b$-quark decay $b \to c + d + \bar u$. 
} 
\label{fig:Tbbud-to-Bm-Dp-pim}
\end{figure}

There are some Feynman diagrams with the help of which the 
three-body purely mesonic final states with the open charm 
in the $T^{\{bb\} -}_{[\bar u \bar d]}$ decays can be identified.   
They are shown in Fig.~\ref{fig:Tbbud-to-Bm-Dp-pim}.   
The factorisable amplitudes of these decays can be written as follows:
\begin{eqnarray} 
&& \hspace*{-5mm} 
{\cal M} ( T^{\{bb\} -}_{[\bar u \bar d]} \to B^- \, D^+ \, \pi^- ) = 
i \, \frac{G_F}{\sqrt 2} \, V_{cb} V_{ud}^* \, a_1^{\rm eff} f_\pi \, p_\pi^\mu 
\label{eq:Tbbud-to-Bm-Dp-pim} \\ 
&& \hspace{3mm} 
\times
\langle (B D)^0_{J_{B D}} (p_{B D}) \left | 
\bar d \gamma_\mu \left ( 1 - \gamma_5 \right ) u 
\right | T^{\{bb\} -}_{[\bar u \bar d]} (p_T) \rangle ,   
\nonumber \\ 
%
&& \hspace*{-5mm} 
{\cal M} ( T^{\{bb\} -}_{[\bar u \bar d]} \to \bar B^0 \, D^0 \, \pi^- ) = 
i \, \frac{G_F}{\sqrt 2} \, V_{cb} V_{ud}^* \, a_1^{\rm eff} f_\pi \, p_\pi^\mu \; 
\label{eq:Tbbud-to-B0-D0-pim} \\ 
&& \hspace{3mm} 
\times 
\langle (B D)^0_{J_{B D}} (p_{B D}) \left | 
\bar s \gamma_\mu \left ( 1 - \gamma_5 \right ) c 
\right | T^{\{bb\} -}_{[\bar u \bar d]} (p_T) \rangle ,   
\nonumber 
\end{eqnarray}
where $a_1^{\rm eff} = C_1 + C_2/N_c$, with $N_c = 3$ being the number 
of quark colors, and the standard definition of the $\pi$-meson leptonic 
decay constant is used: 
\begin{equation} 
\left\langle \pi^- (p_\pi) \left | 
\bar d (0) \gamma^\mu \left ( 1 - \gamma_5 \right ) u (0)
\right | 0 \right\rangle = i f_\pi \, p_\pi^\mu .    
\label{eq:f-pi-definition} 
\end{equation}
Next, we need to parametrise the transition matrix elements from 
the tetraquark state to the double-meson one, having the total angular 
momentum~$J_{B D}$ and zero electric charge (specified by the superscript index), 
in terms of the form factors. 

For the case of the $D$- and $B$-mesons, which are pseudoscalar particles, 
the total momentum~$J_{B D}$ is completely determined by the angular 
momentum~$L_{B D}$ of the system. For the $B D$-system in the $S$-wave, 
the transition matrix elements: 
\begin{eqnarray} 
&& 
\langle (B D)^0_0 (p_{B D}) \left | 
\bar c (0) \gamma_\mu b (0)
\right | T^{\{bb\} -}_{[\bar u \bar d]} (p_T) \rangle ,     
\label{eq:T-to-BD-vector-TME} \\ 
&& 
\langle (B D)^0_0 (p_{B D}) \left | 
\bar c (0) \gamma_\mu \gamma_5 b (0)
\right | T^{\{bb\} -}_{[\bar u \bar d]} (p_T) \rangle ,     
\label{eq:T-to-BD-axial-vector-TME}  
\end{eqnarray}     
define the $S$-wave generalised form factors. 

One can also expect a production of $(B^* D)^0_{J_{B D}}$
or $(B D^*)^0_{J_{B D}}$ pairs in the $S$-wave and the total 
angular momentum ($J_{B^* D} = 1$ or $J_{B D^*} = 1$) is 
determined by the spin of the vector meson ($S_{B^*} = 1$ 
or $S_{D^*} = 1$). This requires another set of $S$-wave 
generalised form factors.

Currently, we lack a reliable dynamical approach to calculate 
the generalised form factors, which prevents us to reliably predict 
the branching fractions for $T^{\{bb\} -}_{[\bar u \bar d]} \to B^- \, D^+ \, \pi^-$ 
and $T^{\{bb\} -}_{[\bar u \bar d]} \to \bar B^0 \, D^0 \, \pi^-$ decays.  
However, as after the $b$-quark weak decay, the emitted charmed quark 
and three spectators have an invariant mass larger than the $B^- D^+$ and $\bar B^0 D^0$ 
thresholds, we expect the split into two hadrons will not cause any dynamical suppression. 
In this case, the transition $T^{\{bb\} -}_{[\bar u \bar d]} \to (B^- \, D^+, \, \bar B^0 \, D^0)$ 
is expected to be comparable with the $B \to D$ transition. 
Accordingly, their branching fractions can reach the value of~$10^{-3}$ as well.

From Fig.~\ref{fig:Tbbud-to-Bm-Dp-pim}, one may speculate that the final $B^- D^+$ and $\bar B^0 D^0$ mesons 
arise from an intermediate bottom-charmed tetraquark state $T^{bc}_{\bar u \bar d}$. 
In the case of a scalar ($J^P = 0^+$) tetraquark, the transition amplitude is governed by the four form factors:  
\begin{eqnarray}
&&  
\langle T^{bc}_{\bar u \bar d} (p_{T'}) | \bar c \gamma_\mu b | T^{\{bb\}}_{[\bar u \bar d]} (p_T) \rangle
= \frac{-2 V (q^2)}{m_T + m_{T'}} \epsilon_{\mu\nu\rho\sigma} \varepsilon_T^\nu p_T^\rho p_{T'}^\sigma , 
\nonumber\\
&&   
\langle T^{bc}_{\bar u \bar d} (p_{T'}) | \bar c \gamma^\mu \gamma_5 b | T^{\{bb\}}_{[\bar u \bar d]} (p_T) \rangle
= 2 i \, m_T \, A_0 (q^2) \, \frac{\varepsilon_T \cdot q}{q^2} \, q^\mu 
\nonumber\\
&&  
+ i \left ( m_T + m_{T'} \right ) A_1 (q^2) 
    \left [ \varepsilon_T^\mu - \frac{\varepsilon_T \cdot q}{q^2} \, q^\mu \right ] 
\nonumber\\
&&   
- \frac{i A_2 (q^2) \left ( \varepsilon_T \cdot q \right )}{m_T + m_{T'}}
     \left[ ( p_T^\mu + p_{T'}^\mu ) - \frac{m_T^2 - m_{T'}^2}{q^2} \, q^\mu \right] ,  
\label{eq:FF-Tbbud-to-Tbcud-0p}
 \end{eqnarray}
where $q^\mu = p_T^\mu - p_{T'}^\mu$, $\varepsilon_T^\mu$ is the polarization vector 
of the axial-vector $T^{\{bb\}}_{[\bar u \bar d]}$ tetraquark, and~$m_T$ and~$m_{T'}$ 
are the masses of $T^{\{bb\}}_{[\bar u \bar d]}$ and $T^{bc}_{\bar u \bar d}$, respectively.  
The decay width for the $T^{\{bb\}}_{[\bar u \bar d]} \to T^{bc}_{\bar u \bar d} \pi^-$ channel is then evaluated as 
\begin{eqnarray}
\Gamma (T^{\{bb\}}_{[\bar u \bar d]} \to T^{bc}_{\bar u \bar d} \, \pi^-) & = & 
2 \times \frac{1}{3} \, \frac{|\vec p_T|}{8 \pi m_T^2} \sum_\lambda \left | {\cal M} \right |^2 , 
\label{eq:Gamma-Tbbud-to-Tbcud-pi}
\end{eqnarray} 
with the decay amplitude 
\begin{eqnarray}
{\cal M} = i \, \frac{G_F}{\sqrt 2} \, V_{cb} V_{ud}^* \, a_1^{\rm eff} f_\pi \times 2 i \, m_T \, A_0 (q^2=0) \left ( \varepsilon_T \cdot q \right ) . 
\label{eq:ME--Tbbud-to-Tbcud-pi}
\end{eqnarray}
The factor~$2$ arises since there are two $b$-quarks in the initial state, while the factor~$1/3$ denotes the spin average. 
Neglecting the $\pi$-meson mass, the decay width can be written as follows: 
\begin{eqnarray} 
\Gamma (T^{\{bb\}}_{[\bar u \bar d]} \to T^{bc}_{\bar u \bar d} \, \pi^-) = 
\frac{G_F^2 m_T^3}{48 \pi} \left | V_{ud} V_{cb}^* \right |^2 f_\pi^2 
\nonumber \\ 
\times 
\left ( a_1^{\rm eff} \right )^2 
\left [ 1 - \frac{m_{T'}^2}{m_T^2} \right ]^3 \left | A_0 (q^2=0) \right |^2 ,  
\label{eq:Gamma-Tbbud-to-Tbcud-pi-exact}
\end{eqnarray} 
where $G_F = 1.166 \times 10^{-5}$~GeV$^{-2}$, $f_\pi \simeq 130$~MeV, $V_{ud} = 0.974$, and $\left | V_{cb} \right | = 40.5 \times 10^{-3}$~\cite{Patrignani:2016xqp}. 
The effective coefficient~$a_1^{\rm eff}$ is a scale-dependent quantity but in a wide range of the energy scale its value is close to unity, so we take $a_1^{\rm eff} = 1$ in estimates.  
Using $m_T = 10.5$~GeV, the $\Xi_{bb} \to \Xi_{bc}$ transition form factor, $A_0 (q^2 = 0) = 0.44$~\cite{Wang:2017mqp}, and $m_{T'} = m_T + m_c - m_b \sim 7.2$~GeV, 
we obtain the estimate of the partial decay width:
\begin{eqnarray}
\Gamma (T^{\{bb\}}_{[\bar u \bar d]} \to T^{bc}_{\bar u \bar d} \, \pi^-) \simeq 7.9 \times 10^{-16}~{\rm GeV} \simeq 1.2~{\rm ns}^{-1} , 
\label{eq:Gamma-Tbbud-to-Tbcud-pi-numeric}
\end{eqnarray}
and the branching fraction with account of the lifetime~(\ref{eq:lifetime-Tbbqqp}):
\begin{eqnarray}
{\cal B} (T^{\{bb\}}_{[\bar u \bar d]} \to T^{bc}_{\bar u \bar d} \, \pi^-) \simeq 1.0 \times 10^{-3} . 
\label{eq:Br-Tbbud-to-Tbcud-pi-numeric}
\end{eqnarray}
If one uses the $B_c\to J/\psi$ transition form factor instead, $A_0 (q^2=0) = 0.53$~\cite{Wang:2008xt}, the results can be enhanced by approximately~$50\%$.  
As the intermediate state, $T^{bc}_{\bar u \bar d}$, will subsequently turn into the~$B^- D^+$ and~$\bar B^0 D^0$ states with equal probabilities, and thus we expect 
\begin{eqnarray}
{\cal B} (T^{\{bb\}}_{[\bar u \bar d]} \to B^- D^+ \pi^-) = 
{\cal B} (T^{\{bb\}}_{[\bar u \bar d]} \to \bar B^0 D^0 \pi^-)  
\nonumber \\ 
\simeq 0.5 \times 10^{-3} . 
\end{eqnarray}
\begin{figure}[tb] 
\begin{center}
\includegraphics[width=0.19\textwidth]{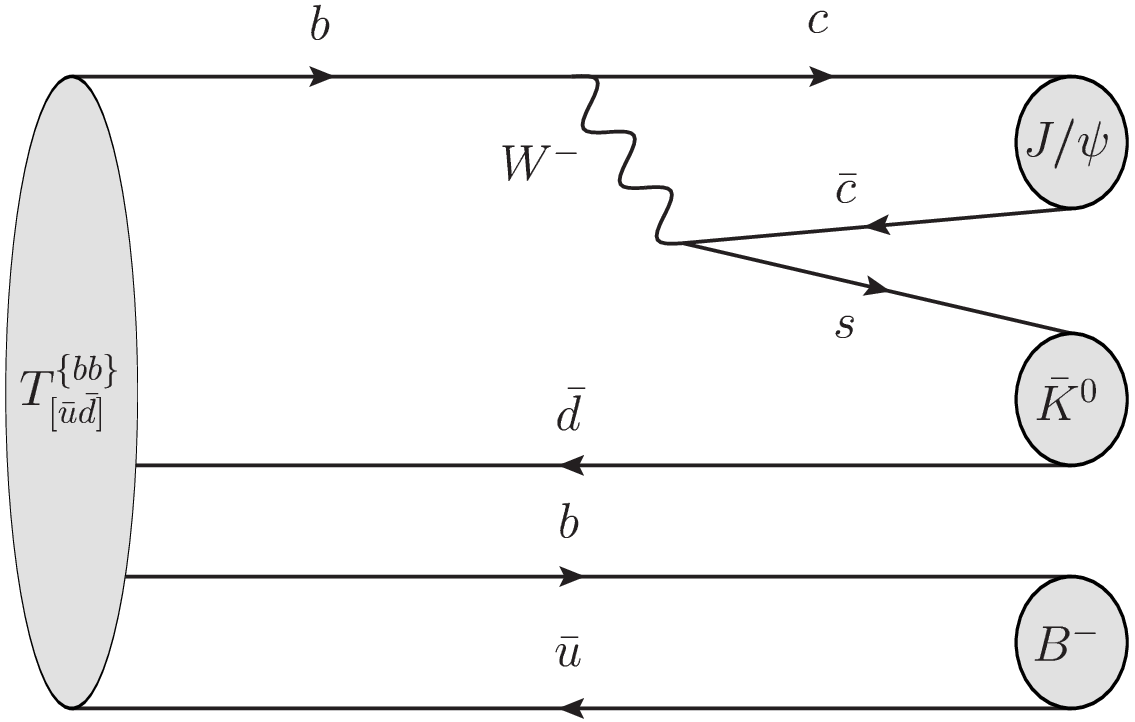}  
\qquad 
\includegraphics[width=0.19\textwidth]{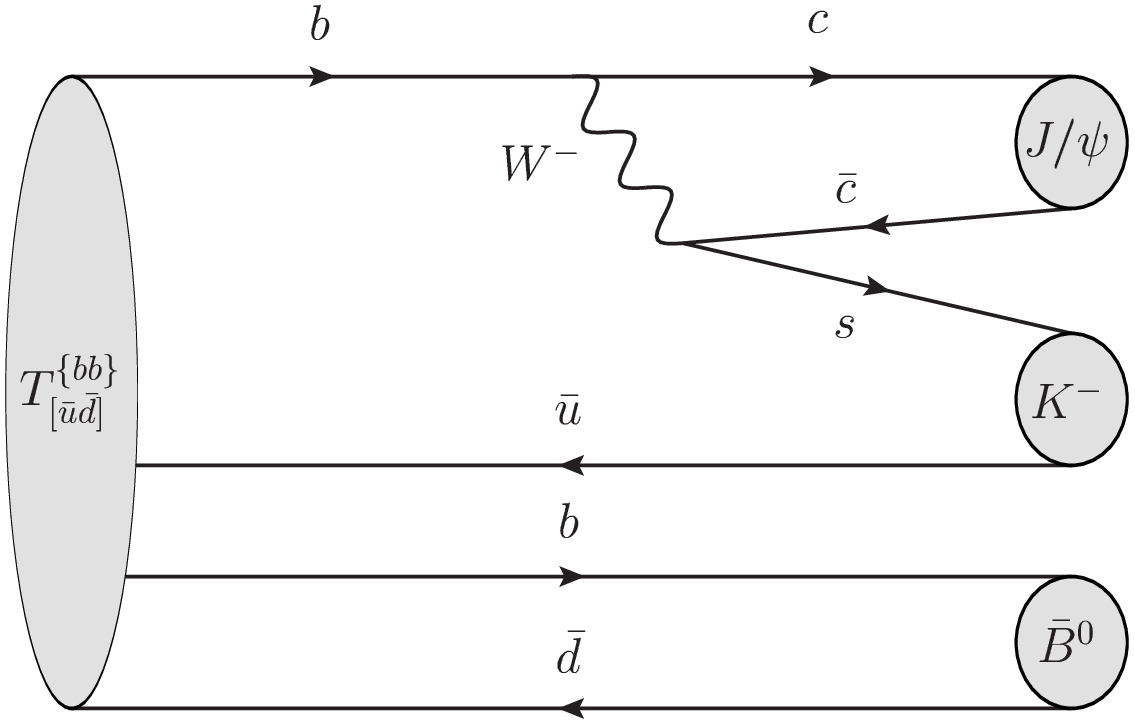}  
\end{center}
\caption{
Feynman diagrams which yield three-body mesonic final states 
in $ T^{\{bb\} -}_{[\bar u \bar d]}$ decays
due to the $b$-quark decay $b \to c + s + \bar c$. 
} 
\label{fig:Tbbud-to-Bm-K0-J-psi}
\end{figure}

\subsubsection{Hidden-Charm Final States} 

There are some Feynman diagrams in which the hidden-charm mesons, such as~$J/\psi$ and~$\psi'$, can be produced.  
The corresponding Feynman diagrams are shown in Fig.~\ref{fig:Tbbud-to-Bm-K0-J-psi}.   
Decay channels include 
\begin{eqnarray}
T^{\{bb\}}_{[\bar u \bar d]} \to J/\psi \overline K^0 B^-, 
\nonumber\\
T^{\{bb\}}_{[\bar u \bar d]} \to J/\psi K^- \overline B^0,  
\label{eq:Tbbud-Jpsi-K-B}
\end{eqnarray}
whose CKM factors are $V_{cb}^* V_{cs}$. 
Their decay branching ratios can be comparable with the $B \to J/\psi K$ decays~\cite{Patrignani:2016xqp}: 
\begin{eqnarray} 
{\cal B} (B^+ \to J/\psi K^+) &=& (1.026 \pm 0.031) \times 10^{-3}, 
\nonumber \\   
{\cal B} (\overline B^0 \to J/\psi \overline K^0) &=& (8.73 \pm 0.32) \times 10^{-4}. 
\end{eqnarray}
So, it reasonable also to expect branching fractions of the $T^{\{bb\}}_{[\bar u \bar d]}$ tetraquark decays~(\ref{eq:Tbbud-Jpsi-K-B}) at the level of few $10^{-4}$.

\subsection{
Three-body baryonic decay modes of $T^{\{bb\} -}_{[\bar u \bar d]}$   
} 

\begin{figure}[tb] 
\begin{center}
\includegraphics[width=0.2\textwidth]{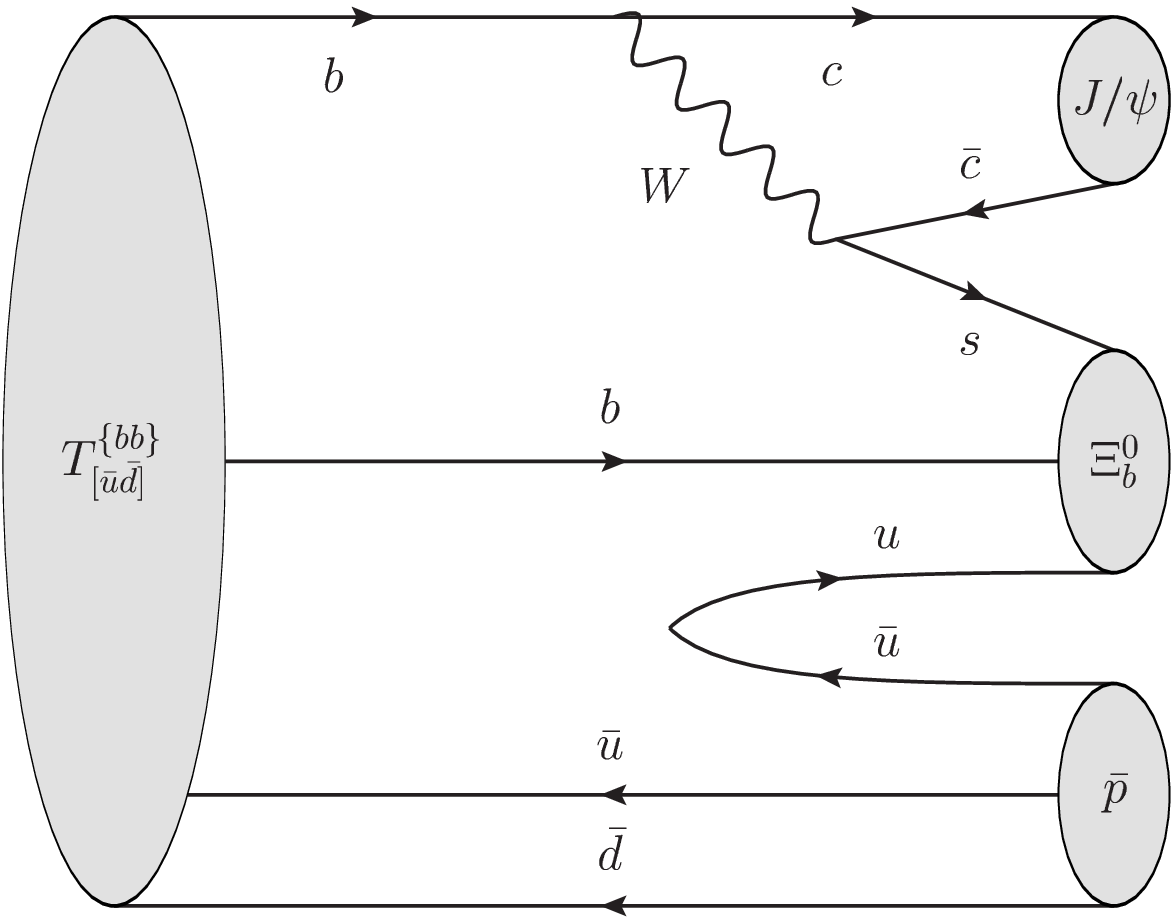} 
\end{center}
\caption{
Feynman diagrams which yield three-body baryonic final states 
in $ T^{\{bb\} -}_{[\bar u \bar d]}$ decays
due to the $b$-quark decay $b \to c + s + \bar c$. 
} 
\label{fig:Tbbud-to-Xib0-p-bar-J-psi}
\end{figure}

The other type of diagrams with hidden-charm have a light- and a $b$-baryon in the final state. 
A representative example of these decays $T^{\{bb\} -}_{[\bar u \bar d]} \to \Xi_b^0 \, \bar p \, J/\psi$,
which can be measured at the colliders, is presented in Fig.~\ref{fig:Tbbud-to-Xib0-p-bar-J-psi}.
As the two-body baryonic decays discussed earlier, the corresponding amplitudes 
are non-factorisable as the quark pair produced in the weak decay is devided and 
hadronised into a baryon and meson. An account of a quark pair picked up from the 
vacuum makes the theoretical analysis even more complicated and we postpone such 
a discussion for future.

\section{Weak Decays of $T^{\{bb\}}_{[\bar u \bar s]} (10643)^-$ and $T^{\{bb\}}_{[\bar d \bar s]} (10643)^0$} 
\label{sec:weak-decays-Tbbus} 

As mentioned earlier, in the case of the double-bottom tetraquarks 
having the strange antiquark\footnote{
The mass assignment is according to the predictions of Ref.~\cite{Eichten:2017ffp}.}, $T^{\{bb\}}_{[\bar u \bar s]} (10643)^-$ 
and $T^{\{bb\}}_{[\bar d \bar s]} (10643)^0$, 
their masses are closer to the corresponding hadronic threshold, $\bar B^* \bar B_s$, 
but are still estimated to lie below it. In particular, these tetraquark states 
are predicted to be below the threshold (10691~MeV)
by 48~MeV in Ref.~\cite{Eichten:2017ffp}. If this is true, their decay modes 
are dominated by flavor-changing charged currents in the effective 
Hamiltonian~(\ref{eq:H-eff-cc}). Like the decays of the tetraquark 
$T^{\{bb\}}_{[\bar u \bar d]}$, they have corresponding 
two-body and three-body decays, which can be 
divided into the purely mesonic and baryonic modes. They are briefly discussed below.

\subsection{Two-Body Baryonic Decays}   
\label{ssec:baryonic-two-body-decays-Tbbus}

As the double-bottom tetraquark $T^{\{bb\}}_{[\bar u \bar s]} (10643)^-$ 
consists of two $b$-quarks and two antiquarks,~$\bar u$ and~$\bar s$, 
the weak decay of one of the $b$-quarks results into two possibilities: 
$b \to c + d + \bar u$ or $b \to c + s + \bar c$. From three quarks 
and three antiquarks one can easily construct baryon and anti-baryon. 
Golden modes are $T^{\{bb\} -}_{[\bar u \bar s]} \to \Xi_{bc}^0 \, \bar\Sigma^-$ 
and $T^{\{bb\} -}_{[\bar u \bar s]} \to \Omega_{bc}^0 \, \bar\Xi_c^-$.  
For $T^{\{bb\}}_{[\bar d \bar s]} (10643)^0$, one can replace the spectator 
$u$-antiquark by the $d$-antiquark which results the decay channels: 
$T^{\{bb\} 0}_{[\bar d \bar s]} \to \Xi_{bc}^0 \left ( \bar\Lambda^0, \, \bar\Sigma^0 \right )$ 
and $T^{\{bb\} 0}_{[\bar d \bar s]} \to \Omega_{bc}^0 \, \bar\Xi_c^0$.   

\subsection{
Three-body Mesonic Decay Modes
}
\label{ssec:3-body-mesonic-decays-Tbbus}

\subsubsection{Open-Charm Final State}

Three-body mesonic final states with the open charm include the following decay modes 
$T^{\{bb\} -}_{[\bar u \bar s]} \to B^- \, D_s^+ \, \pi^-$, $T^{\{bb\} -}_{[\bar u \bar s]} \to \bar B_s^0 \, D^0 \, \pi^-$, 
for the charged strange double-bottom tetraquark and 
$T^{\{bb\} 0}_{[\bar d \bar s]} \to \bar B^0 \, D_s^+ \, \pi^-$, $T^{\{bb\} 0}_{[\bar d \bar s]} \to \bar B_s^0 \, D^+ \, \pi^-$ 
for the neutral one.

\subsubsection{Hidden-Charm Final State} 

There are channels in which the hidden-charm mesons like $J/\psi$, $\psi'$, and etc. can be produced: 
\begin{eqnarray}
T_{[\bar u \bar s]}^{\{bb\} -} \to J/\psi \, \phi \, B^-, \;\;\;
T_{[\bar d \bar s]}^{\{bb\} 0} \to J/\psi \, \phi \, \bar B^0, \nonumber\\
T_{[\bar u \bar s]}^{\{bb\} -} \to J/\psi \, K^- \bar B^0_s,  \;\;\;
T_{[\bar d \bar s]}^{\{bb\} 0} \to J/\psi \, \bar K^0 \, \bar B^0_s. 
\end{eqnarray}
The final states mentioned here are well reconstructed at both electron-positron and hadron colliders.  
%

%

\subsection{
Three-body baryonic decay modes    
} 
\label{ssec:3-body-baryonic-decays-Tbbus}

The other type of diagrams correspond to three-body decays with a light- and bottom baryon in the final state. 
The most interesting processes could be  
$T^{\{bb\} -}_{[\bar u \bar s]} \to \Xi_b^0 \, \bar\Sigma^- \, J/\psi$,  
$T^{\{bb\} -}_{[\bar u \bar s]} \to \Xi_b^- \, \bar\Lambda  \, J/\psi$,  
$T^{\{bb\} 0}_{[\bar d \bar s]} \to \Xi_b^0 \, \bar\Sigma^0 \, J/\psi$ and 
$T^{\{bb\} 0}_{[\bar d \bar s]} \to \Xi_b^- \, \bar\Sigma^+ \, J/\psi$.

\section{$W$-Exchange Diagrams} 
\label{sec:W-exchange}

\begin{figure}[tb]
\begin{center}
\includegraphics[width=0.2\textwidth]{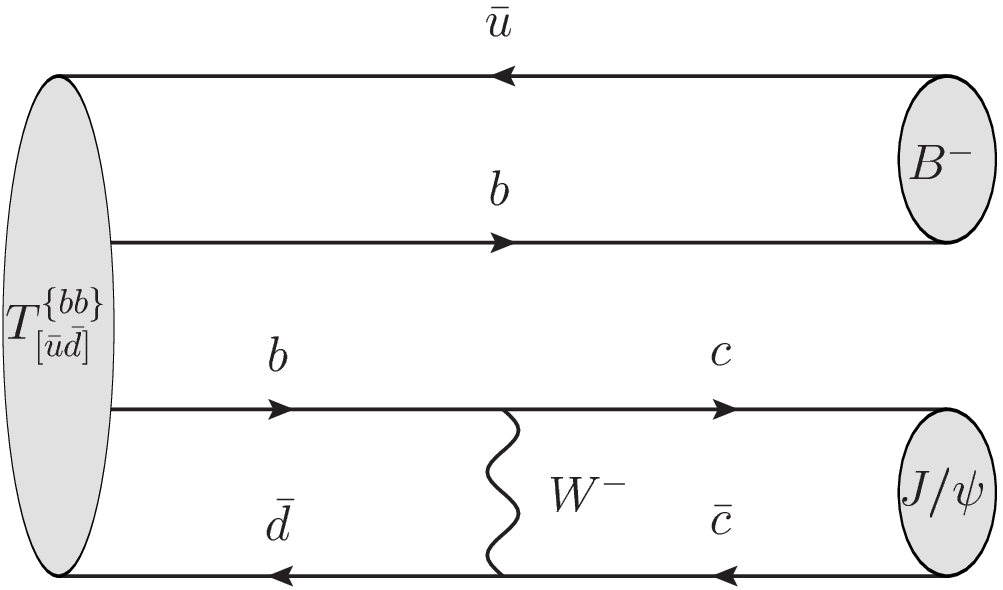} \\ \vspace{0.4cm}
\includegraphics[width=0.2\textwidth]{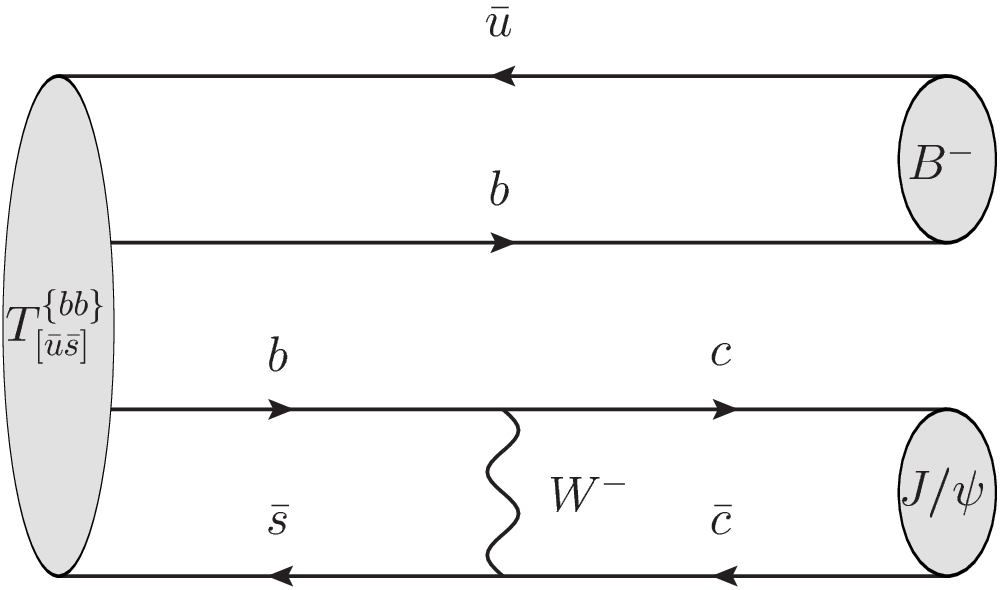} \\ \vspace{0.4cm}
\includegraphics[width=0.2\textwidth]{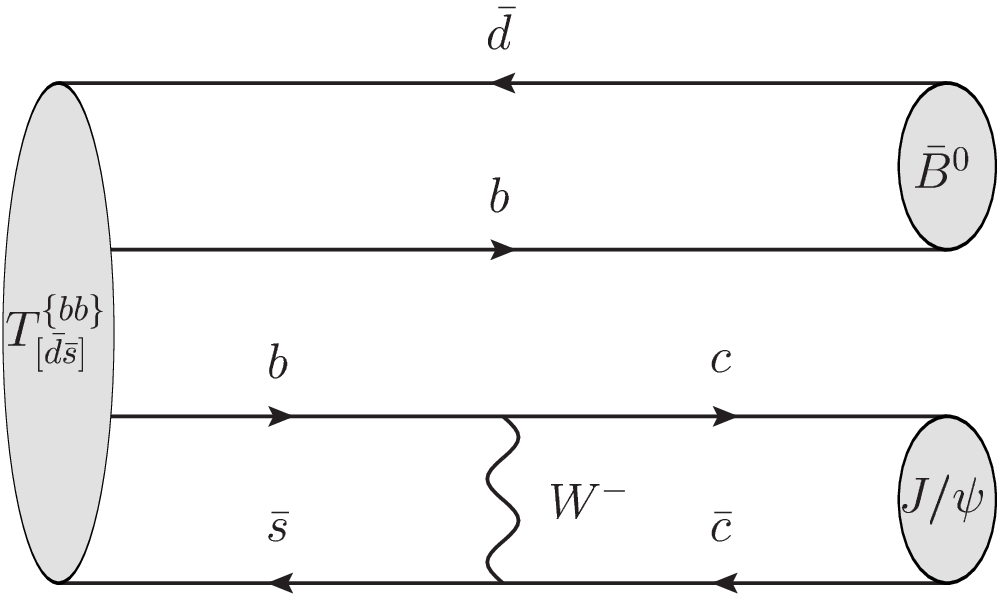} 
\end{center}
\caption{
The $W$-exchange diagrams resulting into two-body mesonic decays 
with the production of the $B$- and $J/\psi$-mesons in $T^{\{bb\}}_{[\bar u \bar d]} $ decays (upper frame),
$T^{\{bb\}}_{[\bar u \bar s]} $ decays (middle frame), and $T^{\{bb\}}_{[\bar d \bar s]} $ decays (lower frame).
} 
\label{fig:Tbbud-to-Bm-J-psi}
\end{figure}

The $W$-exchange diagrams result into two-body mesonic decays, some of which can be of interest at the LHC. 
The use of $W$-exchange decay modes of double-heavy baryons has been advocated in Ref.~\cite{Li:2017ndo}.  
The decays with the $J/\psi$-meson production are the most likely. 
So, the processes which can be searched for are   
$T^{\{bb\}}_{[\bar u \bar d]} (10482)^- \to B^- \, J/\psi$,  
$T^{\{bb\}}_{[\bar u \bar s]} (10643)^- \to B^- \, J/\psi$, and  
$T^{\{bb\}}_{[\bar d \bar s]} (10643)^0 \to \bar B^0 \, J/\psi$. 
The corresponding diagrams for these processes are presented 
in Fig.~\ref{fig:Tbbud-to-Bm-J-psi}.

The decay amplitude to leading order can be factorized. 
For the $T^{\{bb\}}_{[\bar u \bar s]} (10643)^- \to B^- \, J/\psi$ decay, as an example, 
one can write it as follows: 
\begin{eqnarray} 
&&  \hspace*{-3mm} 
{\cal M} ( T^{\{bb\} -}_{[\bar u \bar s]} \to B^- \, J/\psi ) =  
\frac{G_F}{\sqrt 2} \, V_{cd} V_{cs}^* \, a_2^{\rm eff}  
m_{J/\psi} \, f_{J/\psi} \, \varepsilon_\psi^{* \mu} \, 
\nonumber \\ 
&& \hspace*{7mm} \times  
\langle B^- (p_B) \left | \bar s \gamma_\mu \left ( 1 - \gamma_5 \right ) b \right | 
T^{\{bb\} -}_{[\bar u \bar s]} ( p_T, \varepsilon_T ) \rangle , 
\label{eq:M-T-to-Bm-J-psi}  
\end{eqnarray}   
where $a_2^{\rm eff} = C_2 + C_1/N_c$ is the effective Wilson coefficient and the matrix element 
between the vacuum and $J/\psi$-meson is parametrised in terms of the decay constant: 
\begin{equation} 
\langle J/\psi ( p_\psi, \varepsilon_\psi ) \left | 
\bar c  \gamma^\mu P_L c \right | 0 \rangle = 
\frac{1}{2} \, m_{J/\psi} \, f_{J/\psi} \, \varepsilon_\psi^{* \mu} . 
\label{eq:f-J-psi-def}
\end{equation}   
The general decomposition of the transition matrix element can be written 
in the form similar to the $B \to A$ transition, where~$A$ is an axial-vector 
meson~\cite{Li:2009rc}.

An advantage of these decay modes is that there are only two mesons in the final state. 
The $J/\psi$ final state can be easily reconstructed in the $\mu^+\mu^-$ channel, and the bottom meson can be studied in its decays into two-body final states.  
For the $T^{\{bb\}}_{[\bar u \bar s]} (10643)^- \to B^- \, J/\psi$, and $T^{\{bb\}}_{[\bar d \bar s]} (10643)^0 \to \bar B^0 \, J/\psi$, 
the CKM factors in the transition are $V_{cb} V_{cs}^*$, and thus they can have sizeable branching fractions. 
Without any additional power suppression, we expect the branching ratios for these two modes to be of order of~$10^{-3}$.  
For the $T^{\{bb\}}_{[\bar u \bar d]} (10643)^- \to B^- \, J/\psi$ decay, the decay width is suppressed due the CKM factor $V_{cd}$, leading to a smaller branching fraction by approximately the factor~25.

\section{Conclusions}
\label{sec:conclusions}

We have presented the estimate of the double-bottom hadron production in $Z$-boson decays. 
These include the tetraquarks $T^{\{bb\}}_{[\bar u \bar d]}$, $T^{\{bb\}}_{[\bar u \bar s]}$ and $T^{\{bb\}}_{[\bar d \bar s]}$, 
and the double-bottom baryons,~$\Xi_{bb}^0 (bbu)$, $\Xi_{bb}^- (bbd)$, and $\Omega_{bb}^- (bbs)$, which are expected to be produced 
as the jet-fragments of the $(bb)$-diquark jet in the process $ e^+ e^- \to Z \to b \bar b b \bar b$. 
Using the Monte Carlo generators MadGraph5$\_$aMC@NLO and Pythia6, we estimate the branching ratios 
$\mathcal{B} (Z \to  T^{\{bb\}}_{[\bar u \bar d]} + \; \bar b \bar b) = (1.2^{+1.0}_{-0.3}) \times 10^{-6}$, 
and about the same for the sum of the branching ratios $\mathcal{B} (Z \to  T^{\{bb\}}_{[\bar u \bar s]} + \; X)$ and $\mathcal{B} (Z \to T^{\{bb\}}_{[\bar d \bar s]} + \; X)$. 
The summed branching ratios of the double-bottom baryons is also estimated compared to the double-bottom tetraquarks:
$\mathcal{B} (Z \to (\Xi_{bb}^0, \Xi_{bb}^-, \Omega^-_{bb}) + X) : \mathcal{B} (Z \to T^{\{bb\}}_{[\bar q \bar q']} + X) \approx 5.8 : 1$. 
Thus, the double-bottom baryons will be produced about six times more frequently in $Z$-boson decays. 
 
The lifetimes of the tetraquarks $T^{\{bb\}}_{[\bar q \bar q^\prime]}$ are estimated, and for the $SU(3)_F$-multiplet
they are expected to be approximately similar, about one half of the $B$-meson lifetimes, $\tau \simeq 0.8$~ps. 
We also discussed some signature decay modes of these tetraquarks. 
These include the two-body baryonic decays, such as $T^{\{bb\} -}_{[\bar u \bar d]} \to \Xi_{bc}^0 \, \bar p$ and $T^{\{bb\} -}_{[\bar u \bar d]} \to \Omega_{bc}^0 \, \bar \Lambda_c^-$,
whose branching ratios are estimated to be of order of~$10^{-3}$. 
Likewise, three-body mesonic decays of $T^{\{bb\} -}_{[\bar u \bar d]}$ into final states, such as $B^- D^+ \pi^-$ and $B^0 D^0 \pi^-$, are estimated, 
and their branching ratios may also reach order of~$10^{-3}$. 
Of particular interest in this class of decays are those which have hidden-charm mesons in the final states, such as $J/\psi \bar K^0 B^-$ and $ J/\psi K^- \bar B^0$. 
The decay branching ratios ${\cal B} (T^{\{bb\}}_{[\bar u \bar d]} \to J/\psi \bar K^0 B^-)$ and ${\cal B} (T^{\{bb\}}_{[\bar u \bar d]} \to J/\psi K^- \bar B^0)$  
may be comparable to that of $B \to J/\psi K$, which are also of order of~$10^{-3}$~\cite{Patrignani:2016xqp}. 
The decays of the other two members of the $SU(3)_F$-triplet,
$T^{\{bb\} -}_{[\bar u \bar s]}$ and $T^{\{bb\} 0}_{[\bar d \bar s]}$ are also discussed, and the order of magnitude of their signature decay modes are also presented. 
Finally, we emphasise the so-called $W$-exchange diagrams, which give rise to potentially interesting two-body decays $T^{\{bb\}}_{[\bar u \bar d]} (10482)^- \to B^- \, J/\psi$,  
$T^{\{bb\}}_{[\bar u \bar s]} (10643)^- \to B^- \, J/\psi$, and $T^{\{bb\}}_{[\bar d \bar s]} (10643)^0 \to \bar B^0 \, J/\psi$. 
We estimate that the product branching ratios in many of the decay modes discussed here are not expected to exceed $10^{-5}$. 
Hence, several of these decay modes will be required to measure the masses of the tetraquarks to establish them firmly. 
Thus, we recommend to aim for an integrated luminosity of $10^{12}$ $Z$-bosons at the Tera-$Z$ factories being contemplated now. 
They will greatly advance the physics of the double-heavy hadrons, both in the mesonic and baryonic sectors.

\section*{Acknowledgements} 

We would like to thank Marek Karliner, Xiang Liu, Sheldon Stone, Fu-Sheng Yu, Cen Zhang and Zhi-Jie Zhao for helpful discussions. 
We thank Cai-Dian L\"u, the Center for Future High Energy Physics, and IHEP, Beijing, for their generous hospitality.  
This work is supported in part by DFG Forschergruppe FOR 1873 ``Quark Flavour Physics and Effective Field Theories", the National Natural Science Foundation of China under Grant
Nos. 11575110, 11655002, 11735010, and the Natural Science Foundation of Shanghai under Grant No.~15DZ2272100. 
This research is partially supported by the ``YSU Initiative Scientific Research Activity'' (Project No.~AAAA-A16-116070610023-3).

\end{document}